\documentclass[12pt]{iopart}

\usepackage[dvips]{graphicx}
\usepackage{bm}                      
\usepackage{amssymb,pstricks}                 
\begin{document}

\title[Equation of state and compact stars]
{Equation of state at high densities and modern compact star
observations}

\author{David Blaschke}

\address{Institute for Theoretical Physics, University
of Wroclaw, 50-204 Wroclaw, Poland\\
Bogoliubov Laboratory for Theoretical Physics,
JINR Dubna, 141980 Dubna, Russia}
\ead{blaschke@ift.uni.wroc.pl}
\author{Thomas Kl\"ahn}

\address{
Theory Division, Argonne National Laboratory, Argonne IL, USA}
\ead{Thomas.Klaehn@googlemail.com}
\author{Fredrik Sandin}

\address{Department of Physics, Lulea University of
Technology, 97187 Lulea, Sweden}
\ead{Fredrik.Sandin@gmail.com}

\begin{abstract}
Recently, observations of compact stars have provided new data of high
accuracy which put strong constraints on the high-density behaviour of the
equation of state of strongly interacting matter otherwise not accessible in
terrestrial laboratories.
The evidence for neutron stars with high mass ($M =2.1 \pm 0.2 ~M_\odot$
for PSR J0751+1807) and large radii ($R > 12$ km for RX J1856-3754) rules out
soft equations of state and has provoked a debate whether the occurence of
quark matter in compact stars can be excluded as well.
In this contribution it is shown that modern quantum field
theoretical approaches to quark matter including color
superconductivity and a vector meanfield allow a microscopic description of
hybrid stars which fulfill the new, strong constraints.
The deconfinement transition in the resulting stiff hybrid equation of state 
is weakly first order so that signals of it have to be expected due to 
specific changes in transport properties governing the rotational and cooling 
evolution caused by the color superconductivity of quark matter. 
A similar conclusion holds for the investigation of quark deconfinement in
future generations of nucleus-nucleus collision experiments
at low temperatures and high baryon densities such as CBM @ FAIR.
\end{abstract}


\section{Introduction: modern compact star observations}
Compact stars provide natural laboratories for the exploration of baryonic 
matter at high densities, well exceeding in their centres the nuclear 
saturation density of $n_0=0.16$ fm$^{-3}$, where nuclear matter properties 
can be calibrated in terrestrial experiments with atomic nuclei.
Recently, results from observations of compact star properties have been 
reported which have the potential to provide serious constraints for the 
nuclear equation of state (EoS), see \cite{Klahn:2006ir} and references 
therein.
In particular, the high mass of $M=2.1\pm 0.2~M_\odot$ for the pulsar  
J0751+1807 in a neutron star - white dwarf binary system \cite{NiSp05}    
and  the large radius of $R > 12$ km for the   
isolated neutron star RX J1856.5-3754 (shorthand: RX J1856)   
\cite{Trumper:2003we} point to a stiff equation of state   
at high densities.  
Measurements of high masses are also reported for compact stars in low-mass  
X-ray binaries (LMXBs) as, e.g.,  
$M=2.0\pm 0.1~M_\odot$ for the compact object in 4U 1636-536  
\cite{Barret:2005wd}. 
  For another LMXB, EXO 0748-676,   
constraints for the mass $M\ge 2.10\pm 0.28~M_\odot$  
{\it and} the radius $R \ge 13.8 \pm 0.18$ km 
have been reported \cite{Ozel:2006km}.  
The status of these data is, however, unclear since the observation of a   
gravitational redshift $z=0.35$ in the X-ray burst spectra \cite{Cottam:2002}  
could not be confirmed thereafter despite numerous attempts.

Measurements of rotation periods below $\sim 1$ ms as discussed for 
XTE J1739-285 \cite{Kaaret:2006gr}, on the other hand, would disfavor too 
large objects corresponding to a stiff EoS and would thus leave only a tiny 
window of very massive stars in the mass-radius plane 
\cite{Lavagetto:2006ew,Bejger:2006hn}
for a theory of compact star matter to fulfill all 
above mentioned constraints. 
\begin{figure} [ht]
\begin{tabular}{ll}
\includegraphics[angle=270,width=0.5\textwidth]{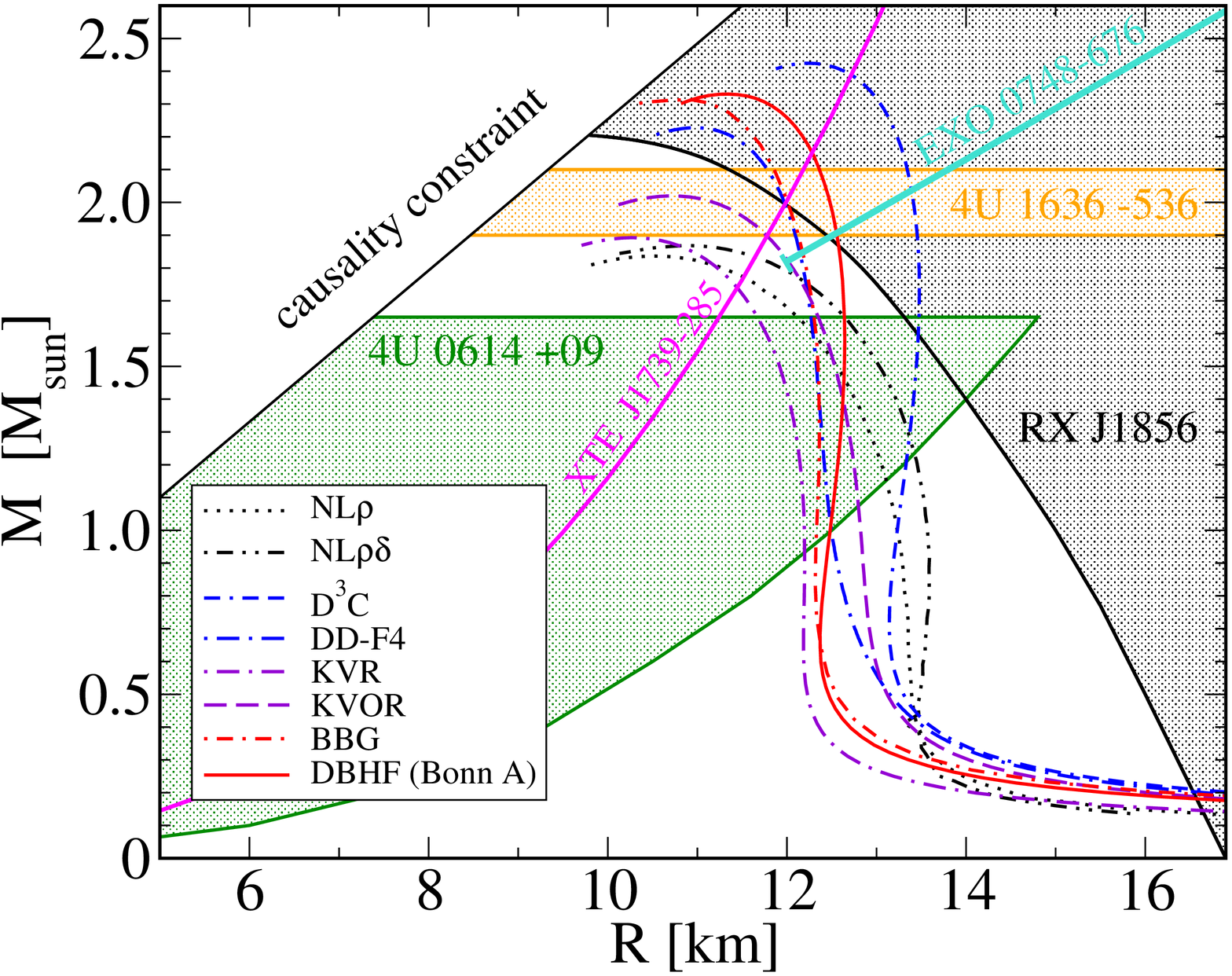}
\includegraphics[angle=270,width=0.5\textwidth]{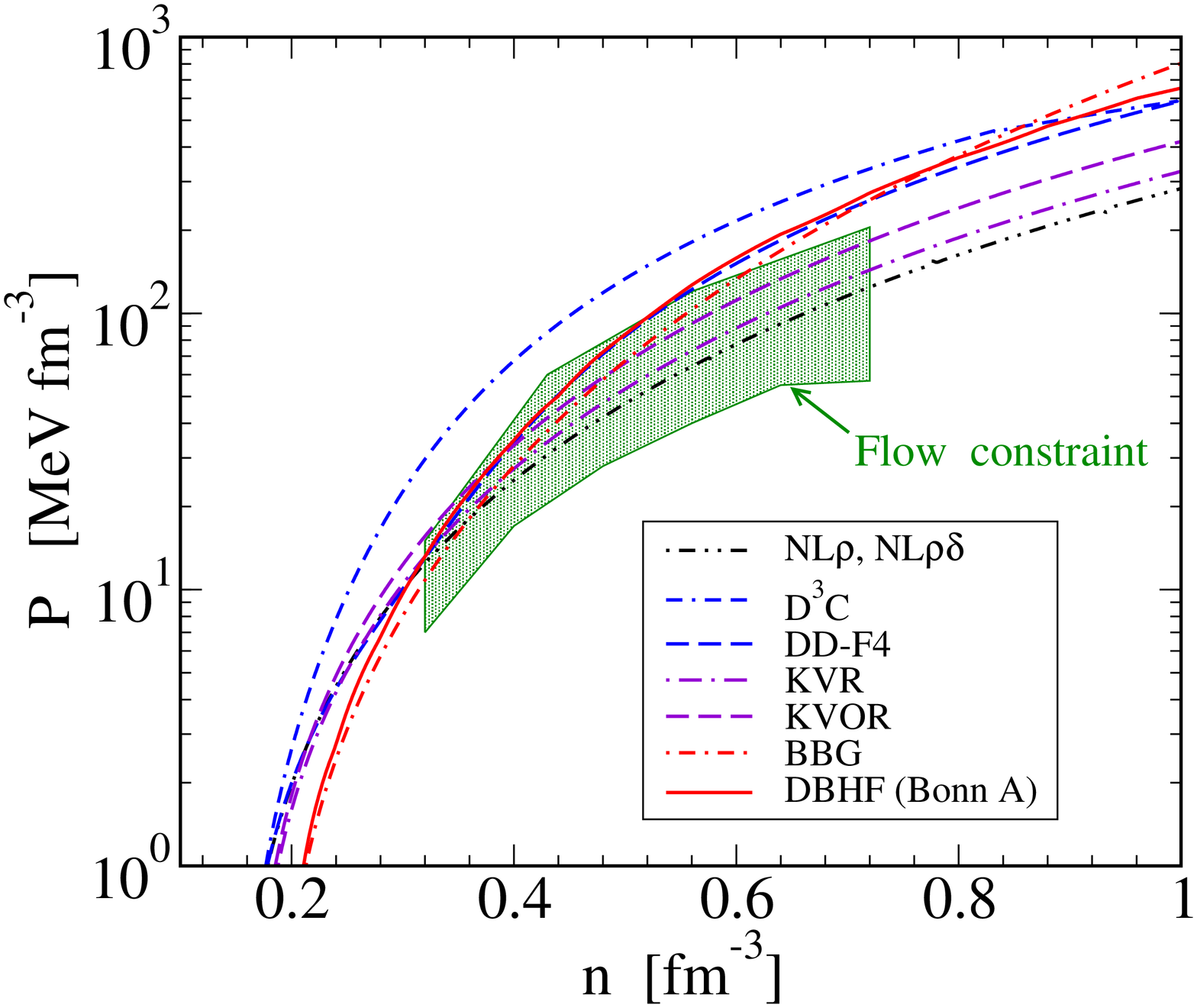}&
\end{tabular}
\caption{Left panel: Mass and mass-radius constraints on compact star 
configurations from recent observations compared to solutions of TOV equations 
for nuclear EoS discussed in the text.  
Right panel: Flow constraint from heavy-ion collisions 
\cite{Danielewicz:2002pu} compared to the same set of nuclear EoS used in the 
left panel. }
    \label{f:M-R}
\end{figure}

In the left panel of Fig.~\ref{f:M-R} we show these modern observational 
constraints for masses and mass-radius relationships together with solutions 
of the Tolman-Oppenheimer-Volkoff (TOV) equations for a set of eight hadronic 
EoS classified in three groups:  
(i) {\it relativistic mean-field (RMF) approaches} 
{\it with non-linear (NL) self-interactions} of the $\sigma$ 
meson \cite{Gaitanos:2003zg}. 
In NL$\rho$ the isovector part of the interaction is
described only by a $\rho$ meson, while the set NL$\rho\delta$ also includes a
scalar isovector meson $\delta$ that is usually neglected in RMF models
\cite{Liu02};
(ii) {\it RMF models with density dependent couplings and 
masses} 
are represented here by four different models from two classes, where in the 
first one density dependent meson couplings are modeled so that a number of  
properties of finite nuclei (binding energies, charge and diffraction radii, 
surface thicknesses, neutron skin in ${}^{208}$Pb, spin-orbit splittings) can 
be fitted \cite{Typel:2005ba}.
D${}^{3}$C has in addition a derivative coupling leading to momentum-dependent
nucleon self-energies and DD-F4 is modeled such that the flow constraint 
\cite{Danielewicz:2002pu} from heavy-ion collisions is fulfilled. 
The second class of these models is motivated by the Brown-Rho scaling 
assumption \cite{Brown:1991kk} that not only the nucleon mass but also the
meson masses should decrease with increasing density. 
In the KVR and KVOR models \cite{Kolomeitsev:2004ff}
these dependences were related to a 
nonlinear scaling function of the  $\sigma$- meson field such that the EoS of 
symmetric nuclear matter and pure neutron matter below four times the
saturation density coincide with those of the Urbana-Argonne group 
\cite{APR}.
In this way the latter approach builds a bridge between the phenomenological 
RMF models and (iii) {\it microscopic EoS} built on realistic 
nucleon-nucleon forces.
Besides the variational approaches (APR \cite{APR}, WFF \cite{Wiringa:1988tp},
FPS \cite{Friedman:1981qw}) such ab-initio approaches to 
nuclear matter are provided, e.g., by the relativistic
Dirac-\-Brueckner-\-Hartree-\-Fock (DBHF) \cite{DaFuFae04} 
and the nonrelativistic Brueckner-\-Bethe-\-Goldstone 
\cite{Baldo:1999rq} approaches.
Stiff EoS like D$^3$C, DD-F4, BBG and DBHF fulfill the demanding constraints
for a large radius and mass, while the softer ones like NL$\rho$ don't. 
It is interesting to note that the flow constraint \cite{Danielewicz:2002pu}
shown in the right panel of  Fig.~\ref{f:M-R} sets limits to the tolerable
stiffness: it excludes the D$^3$C EoS and demonstrates that DD-F4, BBG and DBHF
become too stiff at high densities above $\sim 0.55$ fm$^{-3}$.
For a detailed discussion, see Ref. \cite{Klahn:2006ir}.

A key question asked in investigating the structure of matter at high 
densities is 
whether the phase transition to quark matter can occur inside compact stars. 
In Ref. \cite{Ozel:2006km}, \"Ozel has debated that the new 
constraints reported above would exclude quark matter in compact star 
interiors reasoning that it would entail an intolerable 
softening of the EoS. Alford et al. \cite{Alford:2006vz} have given a few 
counter examples demonstrating that quark matter cannot be excluded.
In the following section we discuss a recently developed chiral quark model 
\cite{Klahn:2006iw} which is in accord with the modern constraints, 
see also \cite{Blaschke:2007ri}.

\section{Phase transition to quark matter: masquerade problem}

The thermodynamics of the deconfined quark matter phase is described  
within a three-flavor quark model of Nambu--Jona-Lasinio (NJL) type, with a 
mean-field thermodynamic potential given by
\begin{eqnarray}  
\Omega_{MF}(T,\mu) 
&=& \frac{1}{8 G_S}\left[\sum_{i=u,d,s}(m^*_i-m_i)^2   
  - \frac{2}{\eta_V}(2\omega_0^2+\phi_0^2) 
        +\frac{2}{\eta_D}\sum_{A=2,5,7}|\Delta_{AA}|^2\right]   
        \nonumber \\  
&&-\int\frac{d^3p}{(2\pi)^3}\sum_{a=1}^{18}  
        \left[E_a+2T\ln\left(1+e^{-E_a/T}\right)\right]  
        + \Omega_l - \Omega_0~.  
\label{eos-blaschke-potential}  
\end{eqnarray}  
Here, $\Omega_l$ is the thermodynamic potential for electrons and muons,  
and the divergent term $\Omega_0$ is subtracted in order to assure zero  
pressure and energy density in vacuum ($T=\mu=0$). The quasiparticle  
dispersion relations, $E_a(p)$, are the eigenvalues of the hermitean matrix  
\begin{equation}  
{\mathcal M} = \left[  
\setlength\arraycolsep{-0.01cm}  
\begin{array}{cc}  
        -\vec{\gamma}\cdot\vec{p}- \hat{m}^*+\gamma^0\hat{\mu}^* &   
        i \gamma_5 C \tau_A \lambda_A \Delta_{AA} \\  
        - C i\gamma_5 \tau_A \lambda_A\Delta_{AA}^* &   
        -\vec{\gamma}^T\cdot\vec{p}+\hat{m}^*-\gamma^0\hat{\mu}^*  
\end{array}  
\right],  
\label{eos-blaschke-eigmatrix}  
\end{equation}  
in color, flavor, Dirac, and Nambu-Gorkov space. Here, $\Delta_{AA}$ are the
diquark gaps. $\hat{m}^*$ is the diagonal renormalized mass matrix and
$\hat{\mu}^*$ the renormalized chemical potential matrix, $\hat{\mu}^*={\rm
  diag}_f
(\mu_u-G_S\eta_V\omega_0,\mu_d-G_S\eta_V\omega_0,\mu_s-G_S\eta_V\phi_0)$.  
The gaps and the renormalized masses are determined by minimization of the
mean-field thermodynamic potential (\ref{eos-blaschke-potential}).  
We have to obey constraints of charge neutrality which depend on the 
application we consider.
In the (approximately) isospin symmetric situation of a heavy-ion collision,
the color charges are neutralized, while the electric charge in general is
non-zero.  For matter in $\beta$-equilibrium in compact stars, also the 
global electric charge neutrality has to be fulfilled. For further details, see
\cite{Blaschke:2005uj,Ruster:2005jc,Abuki:2005ms,Warringa:2005jh}.

We consider $\eta_D$ as a free parameter of the quark matter model, to be
tuned with the present phenomenological constraints on the high-density EoS.
Similarly, the relation between the coupling in the scalar and vector meson
channels, $\eta_V$, is considered as a free parameter of the model.  The
remaining degrees of freedom are fixed according to the NJL model
parameterization in table I of \cite{Grigorian:2006qe}, where a fit to
low-energy phenomenological results has been made.

\begin{figure} [ht]
\begin{tabular}{ll}
\includegraphics[angle=270,width=0.5\textwidth]{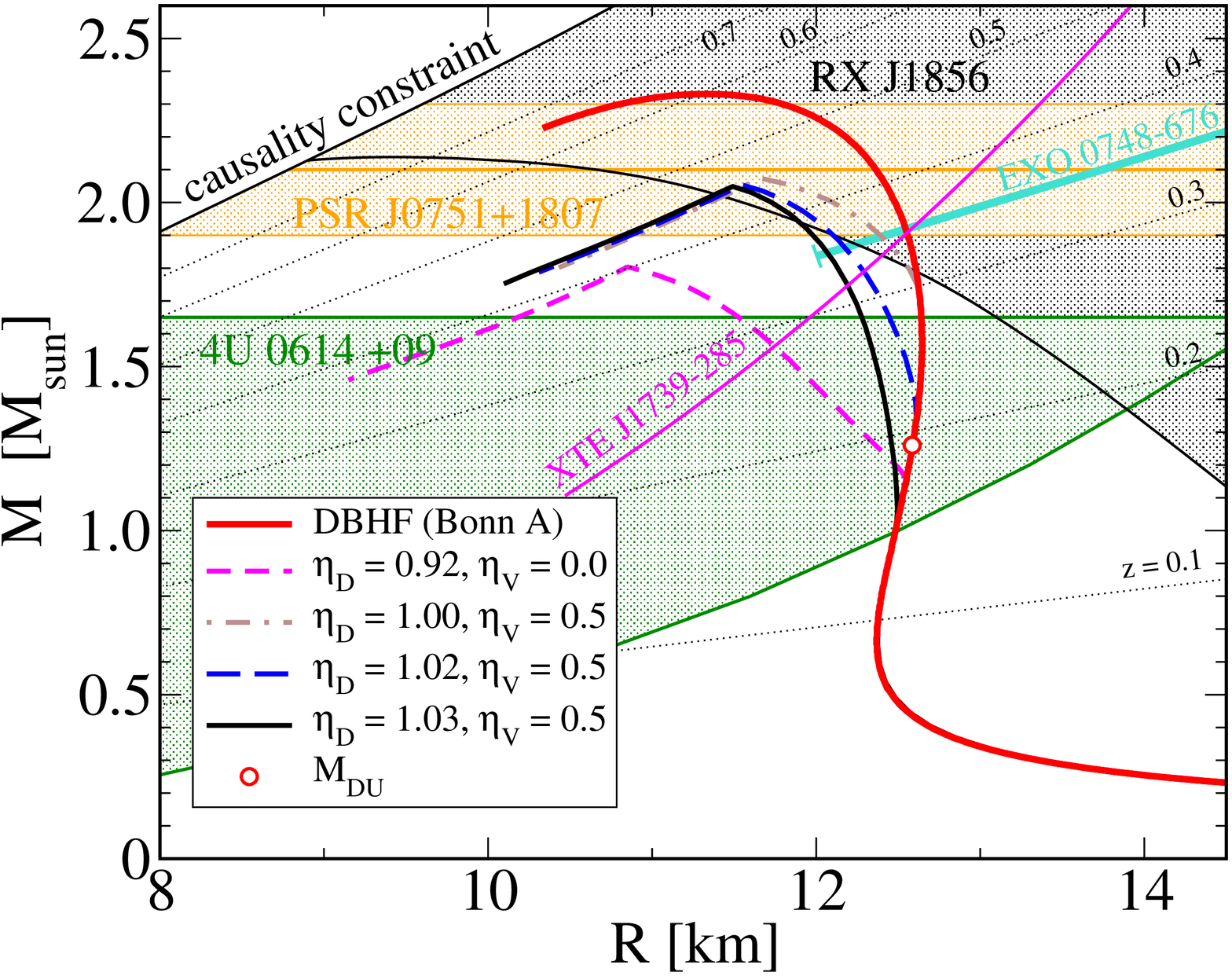}
\includegraphics[angle=270,width=0.5\textwidth]{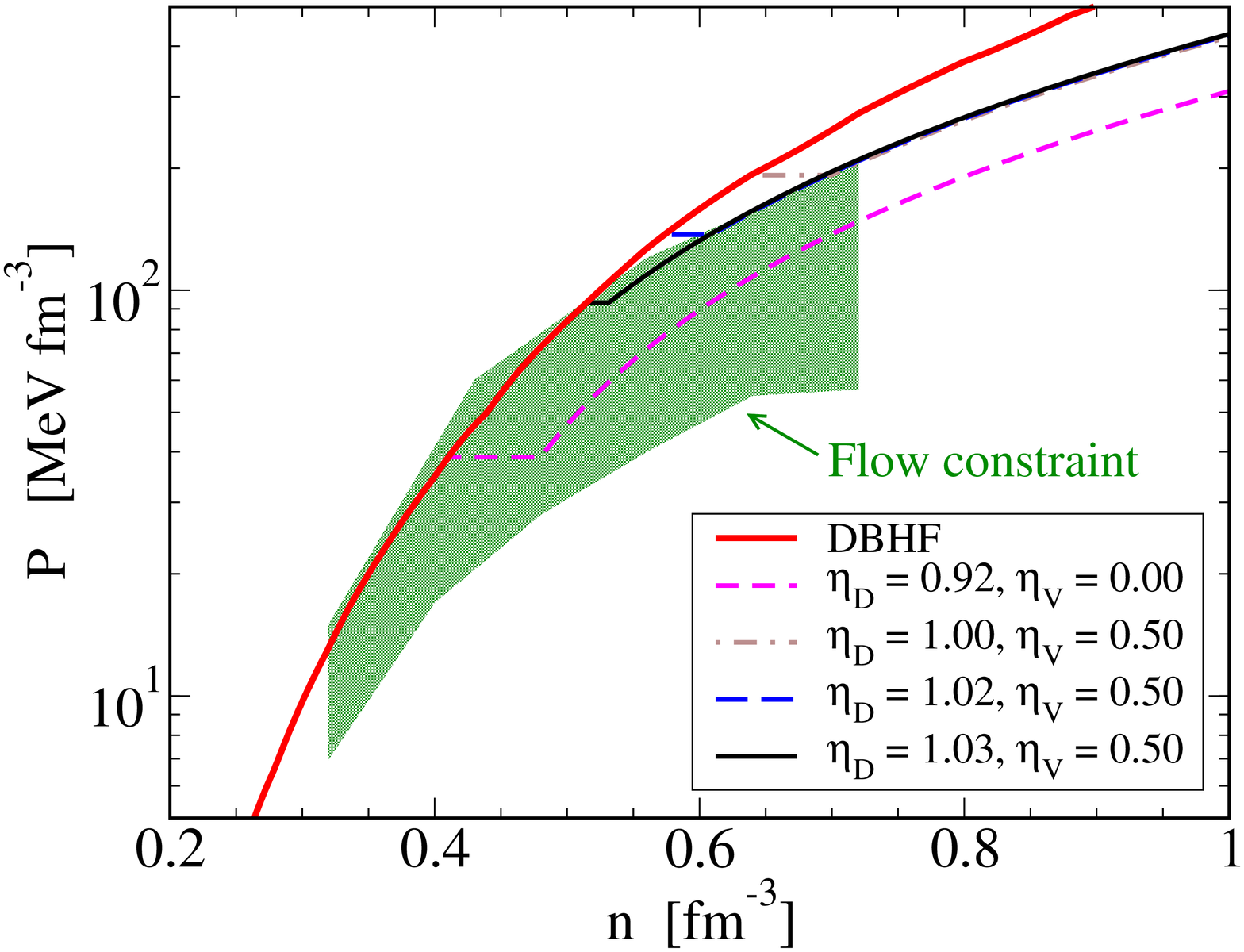}
\end{tabular}
\caption{Same as Fig.~\ref{f:M-R} for hybrid EoS  with a low density hadronic 
branch described by the DBHF approach and a high density quark matter branch 
obtained from a three-flavor NJL model with color superconductivity (diquark 
coupling $\eta_D$) and isoscalar vector meson coupling ($\eta_V$).}
    \label{f:P-n}
\end{figure}

As a relativistic unified description of quark-hadron matter, naturally
including a description of the phase transition, is not available yet, we
apply here the so-called two-phase description, being aware of its
limitations.  For the description of the nuclear matter phase we choose the
DBHF approach and the transition to the quark matter phase given above
is obtained by a Maxwell construction. In the right panel of Fig.~\ref{f:P-n}
it can be seen that the necessary softening of the high density EoS in 
accordance with the flow constraint is obtained for a vector coupling of 
$\eta_V=0.5$ whereas an appropriate deconfinement density is obtained for a 
strong diquark coupling in the range  $\eta_D=1.02 - 1.03$. The resulting phase
transition is weakly first order with an almost negligible density jump.
Applying this hybrid EoS with so defined free parameters under compact star
conditions a sequence of hybrid star configurations is obtained which fulfills
all modern constraints, see the left panel of Fig. 2. 
In that figure we also indicate the minimal mass $M_{DU}$ for which the central
density reaches a value allowing the fast direct Urca (DU) cooling 
process in DBHF neutron star matter to occur, leading to problems with cooling 
phenomenology \cite{Blaschke:2004vq,Blaschke:2006gd}.
Note that for a strong diquark coupling $\eta_D=1.03$, the critical density 
for quark deconfinement is low enough to prevent the hadronic direct Urca (DU) 
cooling problem by an early onset of quark matter. For the given hybrid 
EoS, there is a long sequence of stable hybrid stars with two-flavor 
superconducting (2SC) quark matter, before the occurrence of the strange quark 
flavor and the simultaneous on set of the color-flavor-locking (CFL) phase 
renders the star gravitationally unstable.
Comparing the hybrid star sequences with the purely hadronic DBHF ones one can 
conclude that the former 'masquerade' themselves as neutron stars 
\cite{Alford:2004pf} by having very similar mechanical properties.
To unmask the neutron star interior might therefore require observables 
based on transport properties, strongly modified from normal due to the color
superconductivity.
It has been suggested to base tests of the structure of matter at high 
densities on analyses of the cooling behavior 
\cite{Blaschke:2006gd,Popov:2004ey,Popov:2005xa}
or the stability of fastly rotating stars against r-modes 
\cite{Madsen:1999ci,Drago:2007iy}.
It has turned out that for these phenomena the fine tuning of color 
superconductivity in quark matter is an essential ingredient.

\section{Conclusions}
The evidence for neutron stars with high mass ($M =2.1 \pm 0.2 ~M_\odot$
for PSR J0751+1807) and large radii ($R > 12$ km for RX J1856-3754) rules out
soft equations of state and has provoked a debate whether the occurence of
quark matter in compact stars can be excluded as well.
In this contribution it is shown that modern quantum field
theoretical approaches to quark matter including color
superconductivity and a vector meanfield allow a microscopic description of
hybrid stars which fulfill the new, strong constraints.
The deconfinement transition in the resulting stiff hybrid equation of state 
is weakly first order so that signals of it have to be expected due to 
specific changes in transport properties governing the rotational and cooling 
evolution caused by the color superconductivity of quark matter. 
A similar conclusion holds for the investigation of quark deconfinement in
future generations of nucleus-nucleus collision experiments
at low temperatures and high baryon densities \cite{Grigorian:2006pu},
such as CBM @ FAIR.

\section*{Acknowledgements}
We thank our colleagues who have contributed to the results reported in the 
present work, either with their discussions or by providing us with their 
data. Particular thanks go to M. Alford, F. Burgio, A. Drago, C. Fuchs, 
H. Grigorian, J. Lattimer, M.C. Miller, G. Poghosyan, G. R\"opke, H.J. Schulze,
J. Tr\"umper, S. Typel, D.N. Voskresensky, and F. Weber.
D.B. is supported by the Polish Ministry of Science and Higher 
Education, T.K. is grateful for partial support from GSI Darmstadt and the
Department of Energy, Office of Nuclear Physics, contract no.\ 
DE-AC02-06CH11357.
F.S. acknowledges support from the Swedish Graduate   
School of Space Technology and the Royal Swedish Academy of Sciences.


\section*{References}
\begin{thebibliography}{10}
\bibitem{Klahn:2006ir}     
  Kl\"ahn T {\it et al.} 2006,      
  Phys. Rev. C {\bf 74} 035802.  
  
\bibitem{NiSp05}     
        Nice~D~J, Splaver E M, Stairs I H, L\"ohmer O, Jessner A,     
        Kramer M and Cordes J M 2005,   
        Astrophys. J. {\bf 634} 1242.     
  
\bibitem{Trumper:2003we}  
  Tr\"umper J E, Burwitz V, Haberl F and Zavlin V E 2004,  
  Nucl.\ Phys.\ Proc.\ Suppl.\  {\bf 132} 560.  
  
\bibitem{Barret:2005wd}    
  Barret D, Olive J F and Miller M C 2005,    
  Mon.\ Not.\ Roy.\ Astron.\ Soc.\  {\bf 361} 855.    
       
\bibitem{Ozel:2006km}    
  \"Ozel F 2006, Nature {\bf 441} 1115.    
  
\bibitem{Cottam:2002}  
 Cottam J, Paerels F and Mendez M 2002,   
Nature {\bf 420} 51.  
  

\bibitem{Kaaret:2006gr}
  Kaaret P, {\it et al.} 2006,
Astrophys. J. {\bf 657} L97.

\bibitem{Lavagetto:2006ew}
  Lavagetto G, Bombaci I, D'Ai' A, Vidana I and Robba N R 2006,
  arXiv:astro-ph/0612061.

\bibitem{Bejger:2006hn}
  Bejger M, Haensel P and Zdunik J L 2007,
  Astron.\ Astrophys.\  {\bf 464} L49.

\bibitem{Gaitanos:2003zg}
  Gaitanos T, Di Toro M, Typel S, Baran V, Fuchs C, Greco V and
  Wolter H H 2004,
  Nucl.\ Phys.\ A {\bf 732} 24 (2004).

\bibitem{Liu02} 
  Liu B, Greco V, Baran V, Colonna M and Di Toro M 2002, 
Phys. Rev. C {\bf 65} 045201.

\bibitem{Typel:2005ba}
  S.~Typel,
  Phys. Rev. C {\bf 71}, 064301 (2005).
  
\bibitem{Danielewicz:2002pu}     
  Danielewicz P, Lacey R and Lynch W G 2002,    
  Science {\bf 298} 1592.    

\bibitem{Brown:1991kk}
  Brown G E and Rho M 1991,
  Phys.\ Rev.\ Lett.\  {\bf 66} 2720.

\bibitem{Kolomeitsev:2004ff}  
  Kolomeitsev E E and Voskresensky D N 2005,  
  Nucl. Phys. A {\bf 759} {373}.  
 
\bibitem{APR} 
   Akmal A, Pandharipande V R, Ravenhall D G 1998,
   Phys. Rev. C {\bf 58} 1804.

\bibitem{Wiringa:1988tp}
  Wiringa R B, Fiks V and Fabrocini A 1988,
  Phys.\ Rev.\  C {\bf 38} 1010.



\bibitem{Friedman:1981qw}
  Friedman B and Pandharipande V R 1981,
  Nucl.\ Phys.\  A {\bf 361} 502.

\bibitem{DaFuFae04} 
van~Dalen E N E, Fuchs C and Faessler A 2004,    
Nucl. Phys. A {\bf 744}  227.    
Phys. Rev. C \textbf{72}, 065803 (2005). 

\bibitem{Baldo:1999rq}
  Baldo M, Burgio G F and Schulze H J 2000,
  Phys.\ Rev.\  C {\bf 61} 055801.

\bibitem{Alford:2006vz}   
Alford M, Blaschke D, Drago A, Kl\"ahn T, Pagliara G and Schaffner-Bielich J 
2007,   
Nature {\bf 445} E7.  
  
\bibitem{Klahn:2006iw}
  Kl\"ahn~T {\it et al.} 2006,
  Phys.\ Lett.\ B, in press; [arXiv:nucl-th/0609067].

\bibitem{Blaschke:2007ri}
  Blaschke~D~B, Gomez Dumm~D, Grunfeld~A~G, Kl\"ahn~T and Scoccola~N~N 2007,
  Phys.\ Rev.\  C {\bf 75} 065804.

\bibitem{Blaschke:2005uj}     
  Blaschke D, Fredriksson S, Grigorian H, \"Oztas A M and Sandin F 2005,     
  Phys.\ Rev.\ D {\bf 72} 065020.    
   
\bibitem{Ruster:2005jc}     
  R\"uster S B, Werth V, Buballa M, Shovkovy I A and Rischke D H 2005,     
  Phys.\ Rev.\ D {\bf 72} 034004.     
     
\bibitem{Abuki:2005ms}     
  Abuki H and Kunihiro T 2006,     
  Nucl.\ Phys.\ A {\bf 768} 118.     
  
\bibitem{Warringa:2005jh} 
  Warringa H J, Boer D and Andersen J O 2005, 
  Phys.\ Rev.\  D {\bf 72} 014015. 
 
\bibitem{Grigorian:2006qe}     
  Grigorian H 2007,     
Phys. Part. Nucl. Lett. {\bf 4} 382.  

\bibitem{Blaschke:2004vq}
  Blaschke D, Grigorian H and Voskresensky~D~N 2004,
  Astron.\ Astrophys.\  {\bf 424} 979.

\bibitem{Blaschke:2006gd} 
  Blaschke D and Grigorian H 2007, 
  Prog.\ Part.\ Nucl.\ Phys.\  {\bf 59} 139. 

\bibitem{Alford:2004pf}   
Alford M, Braby M, Paris M W and Reddy S 2005,   
Astrophys.\ J.\  {\bf 629} 969.   
   
\bibitem{Popov:2004ey}     
  Popov~S, Grigorian~H, Turolla~R and Blaschke~D 2006,     
  Astron.\ Astrophys. {\bf 448} 327.     
   
\bibitem{Popov:2005xa}   
 Popov~S, Grigorian~H and Blaschke~D 2006,   
Phys. Rev. C {\bf 74} 025803.  

\bibitem{Madsen:1999ci}
  Madsen J 1999,
  Phys.\ Rev.\ Lett.\  {\bf 85} 10.

\bibitem{Drago:2007iy}
  Drago A, Pagliara G and Parenti I 2007,
  arXiv:0704.1510 [astro-ph].

\bibitem{Grigorian:2006pu}
  Grigorian~H, Blaschke~D and Kl\"ahn~T 2006,
  in: {\it Neutron Stars and Pulsars}, Becker~W and Huang~H~H 
  (eds.), MPE Report {\bf 291} 193; [arXiv:astro-ph/0611595].
\end{thebibliography}
\end{document}